\documentclass[conference]{IEEEtran}
\IEEEoverridecommandlockouts

\usepackage{times}
\usepackage{epsfig}
\usepackage{graphicx}
\usepackage{amsmath}
\usepackage{amssymb}
\usepackage{multirow}
\usepackage{booktabs} 
\usepackage{xcolor}
\usepackage{soul}
\usepackage[pagebackref]
{hyperref}
\usepackage{verbatim}

\begin{document}

\title{Can External Sources Help the Knowledge Cut-off Issues in Audio Deepfake Detection?}

\author{\IEEEauthorblockN{Xin Wang*\thanks{The first three authors contributed equally to this work. XW performed experiments and wrote this paper, JY defined research questions and wrote this paper and XL conducted the initial coding and experiments.}, Junichi Yamagishi*, Xuechen Liu*, Wanying Ge}
\IEEEauthorblockA{National Institute of Informatics, Tokyo, Japan}
}

\maketitle
\begin{abstract}
In conventional deepfake audio detection pipelines, inference and evaluation are typically conducted using standalone, parameterized models trained exclusively on a fixed dataset. In contrast, we investigate a retrieval-augmented detection framework that queries an external knowledge database at inference time—analogous to retrieval-augmented generation in large language models—to mitigate knowledge cutoffs and to find the conditions under which this retrieval-augmented detector outperforms purely parameterized baselines. The external knowledge database can be updated continuously, including during system operation, but its optimal use scenarios have not been previously well characterized. Our empirical results indicate that retrieval augmentation is most beneficial when the content of the knowledge database is distinct from the fine-tuning data of self-supervised learning (SSL)-based models yet remains aligned with the domain of the evaluation set; under these circumstances, linear score-level fusion with standalone detectors is also effective. Nonetheless, the retrieval-augmented system remains vulnerable to domain shifts, and the performance improvements are limited when the test distribution diverges substantially from the domain represented in the knowledge database.
\end{abstract}


\section{Introduction}

\label{sec:intro}
With the advancement of deep neural networks (DNNs), modern audio deepfake detection (ADD) systems, or so-called spoofing {countermeasures} (CM), have achieved strong performance in both controlled and complex scenarios. Such progress is demonstrated  the ASVspoof challenge \cite{asvspoof5}, as well as on recent datasets such as DeepFake-Eval-2024 (DE2024) \cite{deepfakeeval2024}. Meanwhile, recent advances in self-supervised learning (SSL) audio features \cite{baevski2020_wav2vec} have notably improved CM performance on these datasets~\cite{antideepfake}.

In most ADD research, inference and evaluation rely on standalone parameterized detection models trained on datasets of deepfake and genuine audio. When a new deepfake method appears after training, the model lacks information about it, leaving its knowledge effectively cut off.
In addition to such parameterized knowledge-based methodologies, we can also consider a retrieval-augmentation (RA)-based approach that involves preparing an external knowledge database of deepfake and real data, searching for data similar to the input at inference time, and using the information from the retrieved data in the external database to determine whether the input is a deepfake or not. 
Although such methods have already been proposed in the past and their effectiveness has been confirmed~\cite{rad_backend_2024}, the appropriate combinations of external knowledge databases and detection models have not been deeply investigated, and it is not yet properly understood in which scenarios it is optimal to combine external knowledge databases with parameterized detection models.

In this paper, we conduct a series of investigations with the aim of answering the following research questions related to the retrieval-augmented deepfake detection method. 

\noindent
\textbf{RQ1:} When querying an external knowledge database, is it best to use embedding vectors obtained from a parameterized detection model, or should information other than that from the detection model also be taken into account?

\noindent
\textbf{RQ2:} What is the most effective method for making a decision using the information obtained from retrieved samples in the external knowledge database?

\noindent
\textbf{RQ3:} If we fine-tune the detection model on an external knowledge database with a limited set of spoofing methods and thus tailor the model specifically to that database, does the retrieval-augmented approach still offer any benefits?

\noindent
\textbf{RQ4:} Is the retrieval-augmented method using an external knowledge database effective even in cross-database scenarios where the domain of the input data differs from both the training database and the external knowledge database?

\noindent
\textbf{RQ5:} How can we fuse the retrieval-augmented method using an external knowledge database with a standard detector?

While the first two questions are about good configurations of the retrieval-augmented deepfake detection method, the third and fourth questions examine its performance in different evaluation scenarios. 
The last one addresses the synergy between the retrieval-augmented and the standalone detection methods. These research questions, to the best of the authors' knowledge, have not been addressed in any existing study. 

To answer these questions, we propose a simple and comprehensive deepfake audio detection framework utilizing RA from a knowledge database that stores representation-level features and score-level predictions of reference audio files that are extracted by a well-trained SSL-based CM. 
The knowledge database is constructed using a database called DE2024, which contains many latest deepfakes collected from social network platforms~\cite{deepfakeeval2024}. The retrieval method is k-nearest neighbors (k-NN).
To address RQ1, we incorporate speaker and voice-related information, which we call the \textit{voice profile vector}, extracted from specialized foundation models, and investigate whether leveraging such a voice profile matching method can further enhance the retrieval performance.  
To answer RQ2, we compare the following methods on the basis of the scores or labels of similar samples retrieved from the knowledge database: majority voting, calculating the real/fake ratio, and computing the average score. 

In terms of RQ3, we examine a special scenario in which the knowledge database is identical to the fine-tuning data of the standalone detection model and compare the RA's results with those of the standalone detection model fine-tuned on the knowledge database. As for RQ4, we evaluate the cross-domain generalization of the proposed framework on datasets (AI4T~\cite{ai4t} and In-the-wild~\cite{muller22_inthewild}) that are disjoint from the training data and the external knowledge database. Finally, we address RQ5 and compare two training-free fusion strategies. 

Although the proposed framework is applicable to the detection of deepfake audio in general, this study focuses on speech deepfake. 
Related studies are presented in \S~\ref{sec:related}. The proposed RA-based framework is presented in \S~\ref{sec:methods}, covering the external source data (\S~\ref{secsec:source_data}), the construction of the knowledge database (\S~\ref{secsec:knowledge_base}), and the method of retrieving samples and making decisions (\S~\ref{secsec:k-NN}). Experiments addressing RQs 1-3, RQ4, and RQ5 are presented in \S~\ref{sec:indomain}, \S~\ref{sec:crossdatabase_result}, and \S~\ref{sec:fusion}, respectively. The conclusion is drawn in \S~\ref{sec:conclusion}.

\begin{figure*}[t]
    \centering
    \includegraphics[width=0.9\linewidth,trim=5 250 5 5, clip]{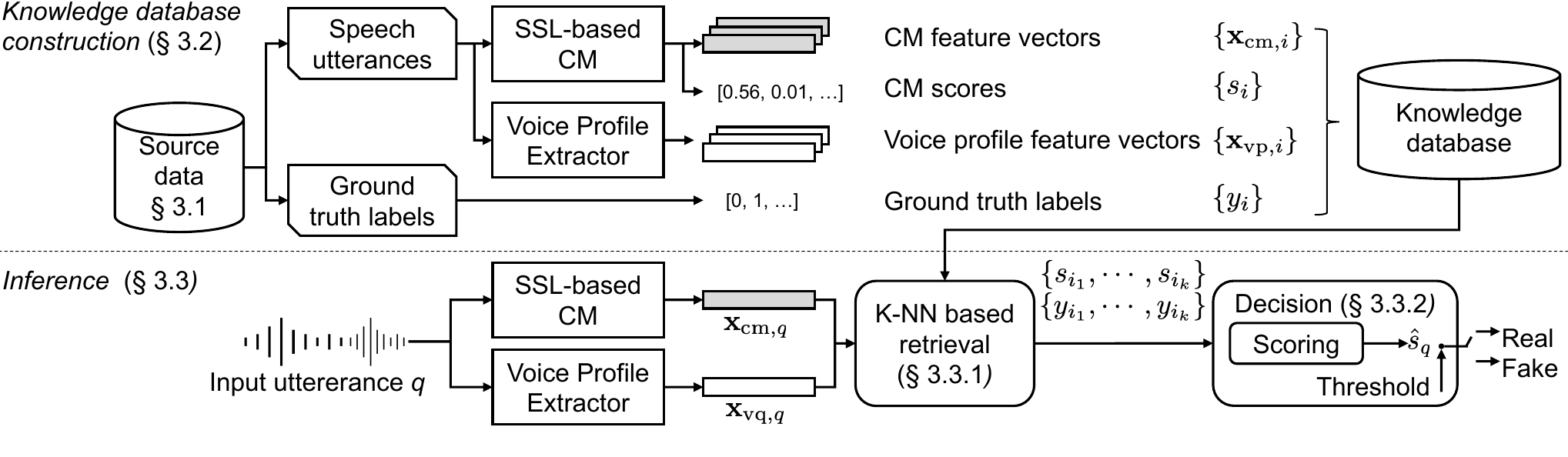}
    \vspace{-5mm}
    \caption{Outline of retrieval-augmented deepfake detection used for our investigation. Knowledge source data contains pairs of speech utterances and their ground truth real/fake labels. 
    Relevant section numbers are marked in figure.
    }
    \vspace{-5mm}
    \label{fig:general}
\end{figure*}

\section{Related Work}
\label{sec:related}
While the ADD field has progressed significantly, the architectural paradigms have remained largely consistent.
The vast majority of existing work relies on standalone parametric approaches, which build binary classifiers, for example, using DNNs such as AASIST~\cite{aasist2022} or SSL-based models~\cite{antideepfake}. The weights of the DNNs are optimized given a fixed training dataset, with the goal of correctly classifying the training data as being either real or fake. During inference, the trained classifier decides between real or fake given an input utterance, and the decision relies entirely on the ``frozen'' knowledge stored within the DNNs' weights. Typical training datasets include those from the ASVspoof challenges~\cite{asvspoof5} and other public sources~\cite{deepfakeeval2024, mlaad}.

A key challenge is whether a model with ``frozen'' knowledge can well detect unseen deepfakes not included in the training set. 
Unfortunately, many existing studies have demonstrated unsatisfying results when standalone detection models have to detect evaluation samples with different recording conditions~\cite{muller22_inthewild}, longer durations~\cite{xuechen_longform2024}, or unseen deepfakes~\cite{mlaad}.

Much like the way of mitigating the knowledge cutoff or hallucination issue when Large Language Models (LLMs) are asked about unknown events~\cite{gao2023retrieval}, one way of improving a deepfake detection model with ``frozen'' knowledge is to augment it with an external library. This concept was recently formalized by Kang et al.~\cite{rad_backend_2024}, who proposed a retrieval-augmented speech deepfake detection framework. Their work constructs a knowledge database that consists of embedding vectors (i.e., query vectors) extracted from speech utterances in an external dataset. The embedding vectors are produced using a standalone detection model. 
At inference time, the nearest vectors are retrieved from the knowledge database. 
The retrieved and the input embedding vectors are then fed into another attention-based detection module, which merges the information and makes the decision between real or fake. 

While our work builds upon RA, we differ by conducting a more granular investigation into the design of the query vectors (RQ1) and different decision strategies (RQ2). 
We also compare the performance of the RA-based method with standalone detection models tuned on an external database (RQ3) and in a more realistic cross-dataset scenario (RQ4). Finally, we investigate the fusion of the RA-based and the standalone deepfake detection methods (RQ5). 

A relevant but conceptually different study has been done for deepfake attribution~\cite{stanTADA2025}. 
The method also retrieves nearest embedding vectors but aims to tell which speech generation algorithm or model was used to forge the input utterance~\cite{stanTADA2025}, given that the input is already known to be fake. 

\section{Methods: Retrieval Augmentation for Speech Deepfake Detection}
\label{sec:methods}

This section describes the investigated retrieval-augmented deepfake detection system. 
A system overview is shown in Fig.~\ref{fig:general}. The upper part illustrates the flow of constructing the knowledge database given source data. The bottom part illustrates the process of retrieval and decision-making when given an input utterance. We explain the source data (\S~\ref{secsec:source_data}), the knowledge database construction (\S~\ref{secsec:knowledge_base}), and the retrieval and decision making (\S~\ref{secsec:k-NN}) in the rest of this section.

\subsection{Source data for knowledge database}
\label{secsec:source_data}

Source data refers to a set of speech utterances to be ``indexed'' and stored in the knowledge database. 
A key assumption in this paper is that \ul{the source data is deliberately constructed to differ from the training data utilized for the deepfake detection model and contains more up-to-date types of deepfakes.} This reflects the real-world disparity between the relatively low cost of generating speech and the much higher cost of training models. Once a deepfake detection system has been trained and deployed, new deepfake generation methods will inevitably emerge while the system is in use. While audio can be generated cheaply and easily, training new detection models requires far more resources and costs. As a result, these new forms of deepfake audio can be collected and added to the external knowledge database with little difficulty, and over time, this external database will increasingly differ from the detection model's training dataset.

In addition, we assume that the source data for the knowledge database includes genuine data from the same domain as the fake data, along with the corresponding ground-truth labels (real/fake). Both of these components are necessary for the decision-making methods examined in this study.  
The data domain is intended to be different from the genuine training data of the deepfake detection model. The motivation is similar to the one for fake data\textemdash collecting diverse genuine data is assumed to be easier than using the data to train the model. 

Accordingly, we use the source data from the training partition of the DE2024 dataset~\cite{deepfakeeval2024}, which contains speech data collected from social media and the latest deepfake detection services. It better represents the latest real-world deepfakes than the training data of the existing standalone detection models, which is described in the next subsection.

\subsection{Knowledge database construction}
\label{secsec:knowledge_base}

The knowledge database is simply a set of entries, where each entry contains the extracted representations and the ground truth label (real/fake) of one speech utterance from the source data. 
Specifically, 
for the entry indexed by $i$, we store the following four types of information.

\noindent 
\textbf{(a) CM feature vector $\mathbf{x}_{\text{cm}, i}$}: We use an open-sourced SSL-based CM model\footnote{\url{https://huggingface.co/nii-yamagishilab/xls-r-2b-anti-deepfake}} to extract a fixed-dimensional embedding vector from the speech utterance in the $i$-th entry. 
This CM uses a Wav2Vec2.0 frontend \cite{baevski2020_wav2vec} and a backend classifier with an adaptive pooling layer, a fully-connected layer, and a softmax output layer. The embedding vector is the output of the pooling layer, which has 1,920 dimensions.

The SSL-based CM model was \emph{post-trained}\footnote{Post-training is an intermediate supervised training phase for SSL models~\cite{antideepfake}, with specifically collected data and training schemes to adapt them for a certain task, which is ADD in this study.} on approximately 74,000 hours of real and fake speech data from various publicly available datasets. \ul{Also note that the post-training datasets do not contain DE2024, and hence, the CM model does not have knowledge of the DE2024 dataset.} 
Further details of model architecture and training are described in \cite{antideepfake}. 
    
\noindent 
\textbf{(b) Voice profile feature vector $\mathbf{x}_{\text{vp}, i}$}: This vector captures multiple voice and speaker attributes using well-trained foundation models~\cite{voxprofile2025}. Various attributes, including age, gender, emotion, and voice quality are estimated. They are then concatenated to form a  profile feature vector. that describes different aspects of the voice and speaker identity. 

Specifically, we use the foundation models from the {Vox-Profile} project\footnote{\url{https://github.com/tiantiaf0627/vox-profile-release}}~\cite{voxprofile2025} to extract four types of attributes: 1) \textit{Age}: a scalar representing the age value in terms of 10 uniformed intervals from 0 to 100 years old\footnote{For example, a value of 7 means an estimated age of 61-70 years old.}; 2) \textit{Gender}: a 2-dimensional one-hot vector for male/female classification; 3) \textit{Emotion}: a scalar indicating the emotion trait (anger, fear, neutral, etc.) and 256-dimensional embedding vector extracted from the intermediate of the foundational model; 4) \textit{Voice quality}: a 25-dimensional vector capturing five perceptual aspects: pitch, volume, clarity, rhythm, and voice texture, following the taxonomy defined in \cite{paraspeechcaps}. 
\ul{Note that the DE2024 dataset is not included in the training data of the pre-trained models that extract the aforementioned attributes}.

\noindent 
\textbf{(d) CM score $s_i \in (0,1)$}: This scalar value indicates how likely the speech in the $i$-th entry is fake. It is set to be the probability of being fake produced by the softmax layer of the same CM that extracts the embedding vector $\mathbf{x}_{\text{cm}, i}$.

\noindent 
\textbf{(c) Ground truth label $y_i \in \{0,1\}$}: This is a ground truth label indicating whether the utterance is real (0) or fake (1). For the DE2024 dataset in our study, $y_i$ is from the manually annotated label included in the original dataset.

\subsection{Retrieval-augmented deepfake detection}
\label{secsec:k-NN}

Using the constructed knowledge database, we perform retrieval-augmented deepfake detection at inference time.
Given a query utterance $q$, we extract the feature vectors and CM score in the same manner as described in the previous subsection. 
Naturally, we lack the true binary label of the query utterance.
We perform k-NN and retrieve the $k$ samples from the knowledge database that are \emph{close} to the query utterance. 
We then use the information from the retrieved samples to determine whether the query is fake or not. The process of retrieval and decision making is illustrated at the bottom of Fig.~\ref{fig:general}. 
Different retrieval and decision-making strategies have been explained in \S~\ref{secsec:retrieval} and \S~\ref{secsec:decision}, respectively. 

\subsubsection{How to retrieve data from knowledge database}
\label{secsec:retrieval}


Three k-NN-based retrieval strategies are investigated in this study. 
\noindent 
\textbf{(a) Retrieval using CM feature vectors only}:  
Let $\mathbf{x}_{\text{cm}, q}$ denote the CM feature vector of the input utterance $q$, which is extracted using the same CM model as that for extracting $\mathbf{x}_{\text{cm}, i}$. 
With the k-NN, the set of retrieved sample indices $\mathcal{N}_{\text{cm}}(q)$ from the knowledge database can be denoted as 
\begin{align}
    &\mathcal{N}_{\text{cm}}(q) = (i_1, \ldots, i_k) \\
    \text{ s.t. } & \mathtt{S}(\mathbf{x}_{\text{cm}, q}, \mathbf{x}_{\text{cm}, i_1}) \geq \ldots \geq \mathtt{S}(\mathbf{x}_{\text{cm},q}, \mathbf{x}_{\text{cm}, i_k}) \\
    \text{and }& \mathtt{S}(\mathbf{x}_{\text{cm}, q}, \mathbf{x}_{\text{cm}, i_k}) \geq \mathtt{S}(\mathbf{x}_{\text{cm},q}, \mathbf{x}_{\text{cm}, j}), \forall j \notin \mathcal{N}_{\text{cm}}(q).
\end{align}
The function $\mathtt{S}(\mathbf{x}_{\text{cm},q},\mathbf{x}_{\text{cm},i}) = \frac{\mathbf{x}_{\text{cm},q} ^\top\mathbf{x}_{\text{cm},i}}{||\mathbf{x}_{\text{cm},q}||_2 \cdot ||\mathbf{x}_{\text{cm},i}||_2},$ measures the cosine similarity between the two vectors
, where $||\mathbf{x}||_2$ denotes the L2 norm of the vector.
Note that the indices in $\mathcal{N}_{\text{cm}}(q)$ are sorted on the basis of similarity. 

\noindent 
\textbf{(b) Retrieval using voice-profile vectors only}: This strategy is the same as the first one except that the nearest indices, denoted as  $\mathcal{N}_{\text{vp}}(q)$, are retrieved on the basis of the cosine similarity between the voice-profile vectors of the query ($\mathbf{x}_{\text{vp}, q}$) and the data in the knowledge database ($\mathbf{x}_{\text{vp}, i}$). 

\noindent 
\textbf{(c) Hybrid retrieval}: The third strategy merges the retrieval outcomes from the first two. It takes the first half of $\mathcal{N}_{\text{cm}}(q)$ and $\mathcal{N}_{\text{vp}}(q)$ and uses their \emph{union} as the retrieved indices $\mathcal{N}_{\text{c+v}}(q)$. 
Note that the size of $\mathcal{N}_{\text{c+v}}(q)$ can be smaller than $|\mathcal{N}_{\text{cm}}(q)| + |\mathcal{N}_{\text{vq}}(q)|$ if there is any index appearing in the first half of both $\mathcal{N}_{\text{cm}}(q)$ and $\mathcal{N}_{\text{vq}}(q)$. 

\subsubsection{How to make a decision using retrieved data}
\label{secsec:decision}
Let $\mathcal{N}(q)$ denote the retrieved indices, which can be either $\mathcal{N}_{\text{cm}}(q), \mathcal{N}_{\text{vp}}(q)$, or $\mathcal{N}_{\text{c+p}}(q)$. 
We derive the score $\hat{s}_q$ for making the final decision using one of the following strategies. 

\noindent 
\textbf{(a) Majority voting:}  $\hat{s}_q$ is either 1 or 0, depending on the majority label among the retrieved data is fake (1) or real (0).

\noindent 
\textbf{(b) Ratio-based score:}  
 $\hat{s}_{q} = \frac{1}{|\mathcal{N}_q|} \sum_{i\in \mathcal{N}_q} \mathbb{I}(y_{i} = 1)$, i.e., the proportion of fake samples out of those retrieved.

\noindent 
\textbf{(c) Average score:} 
    $\hat{s}_{q} = \frac{1}{|\mathcal{N}_q|} \sum_{i\in \mathcal{N}_q} s_i$,
where $s_i$ is the score produced by the CM model.

While the first two strategies require access to the ground truth labels $\{y_i\}$, the last strategy only requires $s_i$. 
The decision is made by comparing $\hat{s}_{q}$ with a decision threshold $\tau$. 
The input is claimed to be \emph{fake} if $\hat{s}_{q}>\tau$, otherwise \emph{real}. This concludes the RA-based inference process.


\section{Experiment I: In-domain Comparison}
\label{sec:indomain}

We present the experiment addressing RQs 1-3 described in \S~\ref{sec:intro}. The experiments addressing the cross-dataset evaluation (RQ4) and the fusion of the RA-based and standalone detection systems (RQ5) are presented in the following sections. 

\subsection{Experimental systems \& data protocols}
Given the three retrieval and three decision strategies, we first build and compare nine RA-based systems (\texttt{S0x}) listed in Table~\ref{tab:main_results}. 
The experiment used the DE2024 dataset. 
Following an existing recipe~\cite{antideepfake}, we segmented the speech utterances in the DE2024 dataset with lengths of either 4, 10, 13, 30, or 50 seconds. 
Experiments were done in each of the five conditions independently. First, we built a knowledge database for the RA-based system using 95\% of the segmented training partition and used the remaining 5\% as a development set for choosing the best number of nearest neighbors $k$, which can be either 5, 10, 20, 50, 100, or 200. Then, the disjoint evaluation partition in the same segmentation-length was used only for evaluation. Note that, although the data for building the knowledge database is from the ``training'' partition of DE2024, RA-based systems require no model training. 

A baseline standalone detection system \texttt{Base} is included for comparison, which is the same as the SSL-based CM model that extracts the CM feature vector in the RA-based systems (\S~\ref{secsec:knowledge_base}~(a)).
This \texttt{Base} uses a publicly available checkpoint, whose training data, albeit on the scale of 74k hours, does not overlap with the knowledge database (i.e., DE2024). 
\texttt{Base} was evaluated using the five segmented versions of the evaluation data.
 
While \texttt{Base} is assumed to have not learned from the latest deepfake data in DE2024, it is useful to investigate the case when the CM model is fine-tuned on the training data of DE2024 to construct a CM model tailored for DE2024. This model may or may not have overwritten knowledge due to catastrophic forgetting caused by fine-tuning. Accordingly, we refer to this system as the topline and denote it as \texttt{Base-FT}. It captures the edge case in which the fine-tuning data of the standalone CM is identical to the source data of the knowledge database (DE2024). 
The fine-tuning uses a learning rate of 1e-7 for ten epochs and other settings consistent with those in~\cite{antideepfake}. 

\begin{table}[!t]
\centering
\setlength{\tabcolsep}{4pt}
\caption{EER (\%) on DE2024 evaluation subsets, grouped by utterance length. $k$ denotes optimal number of retrieved utterances, which was decided based on results on development set. Numbers in \textbf{bold} are lowest in each column of zero-shot evaluation or not significantly different from lowest (at $\alpha=0.05$). Finetuned system is marked by `ft'.}
\begin{tabular}{lrrrrrrrr}
\multicolumn{9}{l}{\textit{Zero-shot evaluation; no fine-tuning with DE2024 training data}} \\
\toprule
& & & & \multicolumn{5}{c}{Evaluation subsets} \\
\cmidrule(lr){5-9}
{ID} & $\mathcal{N}_q$ & $\hat{s}_{q}$ & $k$ & 4s & 10s & 13s & 30s & 50s \\ \midrule
\texttt{S01} & \multirow{3}{*}{\rotatebox{90}{CM}}      & Voting    & 5 & \textbf{15.11} & \textbf{13.45} & \textbf{13.04} & \textbf{11.90} & \textbf{12.15} \\
\texttt{S02} &                                          & Ratio & 5 & \textbf{15.98} & \textbf{14.42} & \textbf{14.27} & \textbf{11.89} & \textbf{12.48} \\
\texttt{S03} &                                          & Avg.   & 10 & 27.22 & 22.11 & 20.55 & 16.71 & 14.92 \\ \midrule
\texttt{S04} & \multirow{3}{*}{\rotatebox{90}{Profile}} & Voting    & 5 & 25.66 & 23.51 & 22.28 & 21.86 & 21.50 \\
\texttt{S05} &                                          & Ratio & 10 & 25.82 & 22.91 & 22.05 & 23.08 & 21.69 \\
\texttt{S06} &                                          & Avg.   & 50 & 33.18 & 29.54 & 27.93 & 31.66 & 30.42 \\ \midrule
\texttt{S07} & \multirow{3}{*}{\rotatebox{90}{Hybrid}}  & Voting    & 20 & \textbf{15.70} & 14.47 & \textbf{13.88} & \textbf{13.08} & 13.87 \\
\texttt{S08} &                                          & Ratio & 5 & 17.42 & 15.06 & 15.16 & 13.68 & \textbf{13.50} \\
\texttt{S09} &                                          & Avg.   & 100 & 27.58 & 22.79 & 20.72 & 23.20 & 20.04 \\ \midrule 
{\texttt{Base}} & -                              & -     & -   & 27.23 & 22.21 & 20.36 & 16.67 & 14.92 \\ \midrule
\multicolumn{9}{l}{\textit{Fine-tuned with DE2024 training data (29 hours)}} \\
\toprule
{\texttt{Base-FT}} & -                              & -     & -   & 12.22 & 11.13 & 10.62 & 9.49 & 8.96 \\ \bottomrule
\end{tabular}
\label{tab:main_results}
\vspace{-5mm}
\end{table}

\label{sec:exp1:result}
\subsection{Results}
We report the equal error rate (EER)\footnote{In this paper we only focuses on how well the scores discriminates fake from real data, hence using EER as the metric~\cite{van2007introduction}. The goodness of score calibration is to be measured in future work.}
in Table \ref{tab:main_results}. \textcolor{black}{A bootstrapping-based method~\cite{Confidence_Intervals} is used to gauge the statistical difference between the EERs of two systems at the significance level of $\alpha=0.05$}. The best values in each column are highlighted in a bold font.  
We summarize the key findings as well as answers to RQs 1-3 in the following paragraphs.

\textbf{Retrieval-augmented deepfake detection approaches vs. parameterized detector.} Comparing the baseline system {\texttt{Base}}, which is post-trained on a large amount of real and spoofed data, with the proposed system that makes decisions using voting and ratio-based scoring, we can see that the retrieval-augmentation method was consistently better than the baseline system {\texttt{Base}}. By leveraging an external knowledge database during inference, it became possible to use information about deepfake techniques that are not included in the post-trained database.

\textbf{RQ1: Query vectors}. In terms of retrieval, using only the profile features (\texttt{S04}-\texttt{S06}) did not yield particularly strong performance. This is likely because the foundation models used for profile feature extraction mainly capture voice and speaker characteristics rather than real/fake cues. However, when these features were combined with the CM features, performance improved.
\textcolor{black}{Combining voice profile and CM features (\texttt{S07}) occasionally produced low EERs, but it is not as good as systems (\texttt{S01} and \texttt{S02}) that only use CM features.}

\textbf{RQ2: Decision making}. Among the three different decision methods, the \textcolor{black}{voting and} ratio-based scoring produced good results. Among the nine variants, including three types of query vectors, the two systems \textcolor{black}{\texttt{S01} and \texttt{S02} achieved the best performance without statistically significant differences.}

\textbf{RQ3: What happens at the edge case in which the fine-tuning dataset for the post-trained SSL-based CM model is perfectly identical to DE2024?} As anticipated, the system {\texttt{Base-FT}} that was directly fine-tuned on a knowledge database having a limited set of spoofing methods, outperformed the retrieval-augmented deepfake detection system with statistically significant differences. We therefore conclude that, when the knowledge database and the fine-tuning training database are exactly the same, using an additional external knowledge database is unnecessary. In other words, the knowledge database only helps when its deepfake methods differ from the information contained in the fine-tuning database. 

These findings are useful not for benchmarks on research databases but for the operational phase, where deepfake detection models are regularly updated. Once a deepfake detection model has been deployed and is in operation, updating the model is costly and can only be done infrequently. \ul{In such scenarios where one aims to enhance detection performance without retraining the detection model, the proposed retrieval-augmented deepfake detection approach proves effective, as it can be easily integrated into any existing systems without any additional training. For instance, when using a 10‑second audio clip as a test sample, simply attaching the retrieval module substantially improves the basic system, {\texttt{Base}}, reducing EER from 22.21\% to 13.45\%.} This performance is slightly inferior to that obtained when the model is updated entirely through fine-tuning, namely {\texttt{Base-FT}} (11.13\%).

Note that in this experiment, because both the fine-tuning set and the evaluation set were drawn from the same dataset, we cannot directly determine whether the tailored systems {\texttt{Base-FT}} and \texttt{S01} are the best or merely overfit the DE2024 dataset. Therefore, the cross-database evaluation in the next section is necessary for further analysis.

\begin{table}[t]
\centering
\caption{Cross-database EER (\%). Numbers in \textbf{bold} are lowest in each column of zero-shot evaluation or not significantly different from lowest (at $\alpha=0.05$). Finetuned system is marked by `ft'. Note that Data size does not cover that for SSL pre-training.}
\setlength{\tabcolsep}{3.5pt}
\begin{tabular}{lrrrrr}
\toprule
& & & \multicolumn{3}{c}{Evaluation data} \\ 
\cmidrule(lr){4-6}
{Model} & SSL type & Data size & {{DE2024(4S)}} & {{AI4T}} & {{ItW}} \\ 
\midrule
\multicolumn{6}{l}{\textit{Zero-shot evaluation, no fine-tuning with DE2024 training data}} \\
\midrule
\texttt{AASIST} & n/a            & 24 hours  & 55.73 & 55.83 & 43.02 \\
\texttt{P3}         & XLSR-53 (0.3B) & 11k hours & 47.83 & 50.96 & 5.87 \\
{\texttt{Base}}    & XLS-R (2B)     & 74k hours & 27.23 & \textbf{10.28} & \textbf{1.23} \\
\texttt{S01}             & XLS-R (2B)    & 74k hours & \textbf{15.11} & 12.93 & 7.76 \\
\midrule
\multicolumn{6}{l}{\textit{Fine-tuned with DE2024 training data (29h)}} \\
\midrule
\texttt{AASIST-FT} & \multirow{4}{*}{\shortstack{Same as \\ above}} & \multirow{4}{*}{\shortstack{fine-tuned \\ on DE2024 \\ 29 hours}} & 18.87 & 53.92 & 16.29 \\
\texttt{P3-FT} & & & 14.04 & 20.36 & 32.82 \\
{\texttt{Base-FT}}    &                                                &                                                                   &  12.22 & \textbf{8.84} & \textbf{4.91} \\
\texttt{S01-FT} &    &      & \textbf{11.82} & 10.87 & 5.56 \\
\bottomrule
\end{tabular}
\label{tab:ood}
\vspace{-5mm}
\end{table}

\section{Experiment II: Cross-database Evaluation}
\label{sec:crossdatabase_result}

\subsection{Experimental systems \& data protocols}

To verify the generalization performance of the RA-based detection method in more general scenarios, we conducted a challenging experiment, where \ul{the evaluation data were selected from an unknown dataset not in the same domain as those of the CM model's training data, fine-tuning data, or external knowledge databases.} We refer to this most challenging scenario as \textit{cross-database evaluation}. 

As the evaluation data in the cross-database scenario, we used the AI4T dataset \cite{ai4t}
and In-the-Wild (ItW) dataset~\cite{muller22_inthewild}, whose deepfake generation methods are significantly different from that of DE2024. The AI4T dataset is a recently released corpus of social media derived speech deepfakes, consisting of 13 hours of audio across eight languages. The In-the-Wild dataset is also a social media derived speech deepfake corpus, but the data was collected in 2022. 

The experimental systems included the best RA-based system identified from addressing RQs 1-2, i.e., \textcolor{black}{\texttt{S01}}. 
It was compared with the baseline \texttt{Base}, as well as another publicly available SSL-based standard detection model (\texttt{P3}~\cite{nii_p3}) and a non-SSLL-based model (\texttt{AASIST}~\cite{aasist2022}), both of which were compared in an existing study using DE2024 4s subset~\cite{deepfakeeval2024}. 

Following the investigation on RQ3, we include the same group of systems but fine-tuned using the corresponding DE2024 training partition. For the RA-based system \textcolor{black}{\texttt{S01-FT}}, its SSL-based front-end for extracting the feature vectors and scores was fine-tuned using the knowledge database.

\subsection{Results}
We list the results for the more challenging cross-database setting in Table~\ref{tab:ood} and present our analysis below. 

\noindent
\textbf{(a) When no fine-tuning with DE2024 training data,} on the DE2024 evaluation data, \textcolor{black}{\texttt{S01}} outperformed \texttt{{Base}} and the strong conventional systems \texttt{AASIST} and \texttt{P3}. However, on the mismatched AI4T and ItW, \textcolor{black}{\texttt{S01}} is inferior to \texttt{{Base}}. This implies that the retrieval-augmented deepfake detection methods suffer from domain mismatch between the knowledge database and the test set. 

\noindent
\textbf{(b) When fine-tuning with DE2024 training data}, 
the tendency was the same as the non-fine-tuning case above: 
 \texttt{S01-FT} performed the best on the DE2024 evaluation data
but was outperformed by {\texttt{Base-FT}} on AI4T and ItW.

These findings suggest that \ul{retrieval-augmented deepfake detection approaches may suffer from domain mismatch issues}, and this serves as the conclusion to \textbf{RQ4}.

\begin{table}[t]
\setlength{\tabcolsep}{3pt}
\centering
\caption{EER (\%) of fusion and non-fusion systems. Numbers in \textbf{bold} are lowest in each column of zero-shot evaluation or not significantly different from lowest (at $\alpha=0.05$). System finetuned on DE2024 (4s) training data is marked by `ft'. }
\begin{tabular}{rrr|rrr}
\toprule
& \multicolumn{2}{c|}{{Evaluation data}} & & \multicolumn{2}{c}{{Evaluation data}} \\
\cmidrule(lr){2-3} \cmidrule(lr){5-6}
Model & DE2024(4s) & AI4T & Fine-tuned {model} & DE2024(4s) & AI4T \\
\midrule
\texttt{Base}                  & 27.23          & \textbf{10.28} & \texttt{Base-FT}                  & 12.22          & \textbf{8.84} \\
\texttt{S01}                   & \textbf{15.11} & 12.93          & \texttt{S01-FT} & \textbf{11.82} & 10.87         \\
\texttt{Base+S01} & 16.90          & \textbf{10.38} & \texttt{(B01+S01)-FT} & \textbf{12.10} & \textbf{8.77} \\
\texttt{Base|S01}  & 18.57          & 19.17          & \texttt{(B01|S01)-FT} & 13.41          & 11.24         \\
\bottomrule
\end{tabular}
\label{tab:fusion}
\vspace{-5mm}
\end{table}

\section{Experiment III: Fusion of RA-based and Standard Detection Systems}
\label{sec:fusion}


The final experiment compares the RA-based system, the conventional detection model, and their fusion. Here, we examine a linear and a selective fusion strategy. The experiments follow the previous ones and investigate the performance with and without fine-tuning on DE2024. The evaluation sets include the evaluation partition of DE2024 and AI4T. 


\noindent
\textbf{(a) {Linear fusion}}: given CM score ($s_{q}$) and the score from the RA-based strategy ($\hat{s}_{q}$), the fused score is set to $\lambda s_{q} + (1 - \lambda)\hat{s}_{q}$. The value of $\lambda$ was set to 0.9. 

\noindent
\textbf{(b) {Selective fusion}}: The query representation is first input to a specialized module for deciding whether it is in-domain or out of domain (OOD) and then routed to the most suitable detector. If the query is judged as in-domain, we directly use the score from the SSL-based CM ($s_{q}$). Conversely, if the query is judged to be OOD, we switch to the retrieval-augmented deepfake detection method with its knowledge database and use the decision score ($\hat{s}_{q}$) as the final output. 
We used a training-free $k$-NN based OOD that measures cosine similarity between an input embedding and one in a knowledge database \cite{ood_knn_density}. 
We utilized two knowledge databases with distinct roles: 1) a knowledge database dedicated to training-free OOD detection, and 2) another database used for the retrieval-augmented deepfake detection system. 
For the $k$-NN-based OOD detection, $k$ was set to 20. 

For the experiment, we fused \textbf\texttt{Base} and \texttt{S01}. The linear and selective fusion methods of two systems are denoted by \texttt{Base+S01} and \texttt{Base|S01} in Table~\ref{tab:fusion}, respectively. Similar notations are used for fusion of 
 fine-tuned models.
%
As the results in Table~\ref{tab:fusion} demonstrate, \textcolor{black}{the linear fusion tends to be effective for combining the retrieval-augmented detector and a conventional detector. It is only inferior to \text{S01} on the DE2024 evaluation data when fusing non-fine-tuned models (\text{Base+S01}).}
This serves as the conclusion to \textbf{RQ5}.

\section{Conclusion}
\label{sec:conclusion}
In this paper, we conducted an in-depth analysis of a retrieval-augmented deepfake detection system that leverages an external knowledge database, and we identified the conditions under which the external knowledge database is beneficial. Although this external knowledge database can be updated even after model training, its optimal usage scenario had been unclear. Our experimental results show that the external knowledge database is most useful when its contents differ from the fine-tuning data of the SSL-based CM and, at the same time, are similar to the domain contained in the test set. Under these conditions, we also found that linear fusion with standalone systems is effective. On the other hand, our results clearly demonstrate that the retrieval-augmented deepfake detection system using the external knowledge database is also strongly affected by domain shifts and that performance improvements are limited when the domain of the test set differs significantly.
%

\vspace{1mm}
\noindent
\footnotesize{\textbf{Acknowledgments}: This study is based on results obtained from a project, JPNP22007, commissioned by the New Energy and Industrial Technology Development Organization (NEDO). This study was carried out using the TSUBAME4.0 supercomputer at Institute of Science Tokyo.}

\bibliographystyle{IEEEtran}
\bibliography{ref}
\end{document}